\documentclass[prl,reprint, twocolumn,showpacs,preprintnumbers,amsmath,amssymb,nofootinbib,floatfix]{revtex4-1} 
\usepackage{graphicx}

\usepackage{mathrsfs}
\usepackage{hyperref}

\usepackage{slashed}

\usepackage{color}

\usepackage{gensymb}

\def \matrix #1 {\left(\begin{array}{cc} #1 \end{array}\right)}

\def\II{\hbox{{1}\kern-.25em\hbox{l}}}

\begin{document}

\title{Enhanced Next-to-Leading-Order Corrections to Weak Annihilation $B$-Meson  Decays}

\author{Cai-Dian L\"{u}$^{a,b}$}
\email{lucd@ihep.ac.cn}

\author{Yue-Long Shen$^{c}$}
\email{corresponding author: shenylmeteor@ouc.edu.cn}

\author{Chao Wang$^{d}$}
\email{corresponding author: chaowang@nankai.edu.cn}

\author{Yu-Ming Wang$^{e}$}
\email{corresponding author: wangyuming@nankai.edu.cn}

\affiliation{${}^a$  Institute of High Energy Physics, CAS, P.O. Box 918(4), Beijing 100049,  China \\
${}^b$   School of Physics, University of Chinese Academy of Sciences, Beijing 100049,  China \\
${}^c$ School of Physics and Optoelectronic Engineering, Ocean University of China,
Songling Road 238, Qingdao, Shandong 266100, P.R. China \\
${}^d$ Department of Mathematics and Physics,
Huaiyin Institute of Technology,
Meicheng East Road 1,  Huaian, Jiangsu 223200, P.R. China \\
${}^e$ School of Physics, Nankai University,
Weijin Road 94, Tianjin 300071, P.R. China}

\date{\today}

\begin{abstract}
\noindent
We accomplish the analytical computation of the  pure weak annihilation
non-leptonic $B$-meson decay amplitudes  at leading power in the heavy quark expansion.
The novel observation regarding such fundamental hadronic quantities
is that adding the missing hard-collinear contribution on top of the  hard gluon exchange effect
eliminates  rapidity divergences entering the  convolution integrals
of factorization formulae.
Subsequently we identify the  perturbative enhancement mechanism due to
the penguin contractions of the current-current operators from the weak effective  Hamiltonian,
which yields the significant impacts on the CP violating observables.
\\[0.4em]

\end{abstract}


\maketitle

%
\section{Introduction}
%

It is generally accepted that the exclusive two-body charmless bottom-meson decays
are of fundamental importance for advancing our understanding
towards diverse facets of the strong interaction dynamics governing
the flavour-changing heavy-quark decay processes and for exploring the
peculiar implications of the Cabibbo-Kobayashi-Maskawa (CKM) mechanism
for CP violation in electroweak interactions.
The primary challenge of predicting such hadronic $B$-meson decay observables
consists in constructing a field-theoretic framework to disentangle
the calculable perturbative QCD fluctuations from the non-perturbative
soft and collinear physics systematically.
In this respect, the QCD factorization formalism based upon expansions in the small parameter
$\Lambda_{\rm QCD} / m_b$ \cite{Beneke:1999br,Beneke:2000ry,Chay:2003ju,Beneke:2002ph,Bauer:2004tj,Bauer:2005kd}
has proven to provide the precise prescriptions for
evaluating the appeared non-leptonic matrix elements of the effective weak Hamiltonian operators.
However, the long-standing obstacle to improve  the factorization calculations
of the two-body $B$-meson decay amplitudes arises from the disturbing end-point divergences
in the convolution integrals describing the power-suppressed
but phenomenologically pronounced weak annihilation corrections.
In particular, the inadequacy of achieving the factorization-compatible regularization
of the annihilation amplitudes 
apparently does not allow  to provide quantitative predictions for the annihilation-dominated decay processes.

Additionally, the abiding lack of higher-order perturbative corrections to
the weak annihilation topologies   prevents us from
obtaining  reliable estimates of the strong-phase sensitive quantities
(for instance, the direct CP asymmetry) due to a variety of potential enhancements,
which do not manifest themselves at  leading-order (LO) in the strong coupling.
The process-independent parametrizations for logarithmically and linearly divergent integrals
in the basic building blocks for the exclusive hadronic $\bar B_q  \to PP$, $\bar B_q  \to PV$
and $\bar B_q  \to VV$ (with $q=d, \, s$) decay amplitudes \cite{Beneke:2001ev,Beneke:2003zv,Beneke:2006hg,Bartsch:2008ps}
(see also \cite{Cheng:2008gxa,Cheng:2009cn,Cheng:2009mu} and references therein) introduce further assumption
for the current QCD factorization  predictions,
especially with the finite bottom-quark mass in practice,
which has been examined exploratorily by employing the data-driven strategy
\cite{Zhu:2011mm,Wang:2013fya,Bobeth:2014rra,Chang:2014yma,Chang:2016qyc,Chang:2017brr}.
The striking importance of the weak annihilation mechanism
in investigating the non-leptonic $B$-meson decay amplitudes
also motivated the interesting dynamical anatomy \cite{Arnesen:2006vb,Khodjamirian:2005wn,Feldmann:2004mg}
on the basis of distinct QCD techniques
(see \cite{Keum:2000wi,Lu:2000em,Li:2004ep,Ali:2007ff,Li:2010nn,Li:2012nk,Li:2012md,Li:2013xna,Li:2014xda}
for an alternative treatment taking into account the non-vanishing transverse momenta of the active partons).
As a consequence, accomplishing the systematic computation of the weak annihilation topologies
with the non-trivial  strong phases will be of the top priority for developing the conceptual framework
of the charmless two-body $\bar B_q  \to M_1 M_2$ decays and for better confronting the anticipated precision measurements
of a large number of CP violating observables at the Belle II experiment \cite{Belle-II:2018jsg}.

In order to probe factorization properties of the weak annihilation contributions immaculately,
we report on a novel observation of the hard-collinear gluon exchange
(the short-distance configuration with the invariant mass of ${\cal O} (\sqrt{m_b \, \Lambda_{\rm QCD}}))$
in regularizing  the unwanted rapidity divergences analytically,
which has been unfortunately neglected in all previous calculations of the two-body hadronic $B$-meson decays,
by concentrating on the pure annihilation channels such as $\bar B_s \to \pi \pi$ and  $\bar B_d \to \phi \phi$.
The significance of this  essential  insight  for consolidating the theory foundation of QCD factorization
in $B$-meson decaying into two light mesons will be further strengthened by inspecting
a sample set of the next-to-leading order (NLO) Feynman diagrams, which are enhanced by the large Wilson coefficients
and/or the  multiplication  CKM factors numerically
and are promising to  yield the sizeable strong phases.
General implications of the improved formalism on the accessible decay observables
will be then discussed with the acceptable models for the leading-twist
light-cone distribution amplitudes  of both the bottom meson and the final-state hadrons.

%
\section{General analysis}

We start by setting up our notation of the effective weak Hamiltonian
of the non-leptonic $\Delta B =1$ transitions in the Standard Model (SM)
\begin{eqnarray}
{\cal H}_{\rm eff} &=& {4 \, G_F \over \sqrt{2}} \,
\sum_{p=u, c} V_{p b} \, V_{p q}^{\ast}
\bigg [ C_1(\nu) \, P_1^{p}(\nu)  + C_2(\nu) \, P_2^{p}(\nu)
\nonumber \\
&& + \sum_{i=3}^{6} C_i(\nu) \, P_i(\nu)
+ \sum_{i=3}^{6} C_{i Q}(\nu) \, P_{i Q}(\nu)
\nonumber \\
&& + C_{7 \gamma}(\nu) \,\, P_{7 \gamma}(\nu)
+ C_{8 g}(\nu) \,\, P_{8 g}(\nu)   \bigg  ] \,,
\end{eqnarray}
where we adopt the effective operator basis introduced in
\cite{Chetyrkin:1996vx,Chetyrkin:1997gb,Bobeth:1999mk,Bobeth:2003at,Huber:2005ig}
enabling the disappearance of Dirac traces with $\gamma_5$.
The renormalized Wilson coefficients $C_i(\nu)$ with $\nu \sim {\cal O}(m_b)$ will be evaluated  for our purpose
in the next-to-leading-logarithmic (NLL) approximation  in the Chetyrkin-Misiak-M\"{u}nz  basis
\cite{Chetyrkin:1997gb}.

\begin{figure}[tp]
\begin{center}
\includegraphics[width=1.0 \columnwidth]{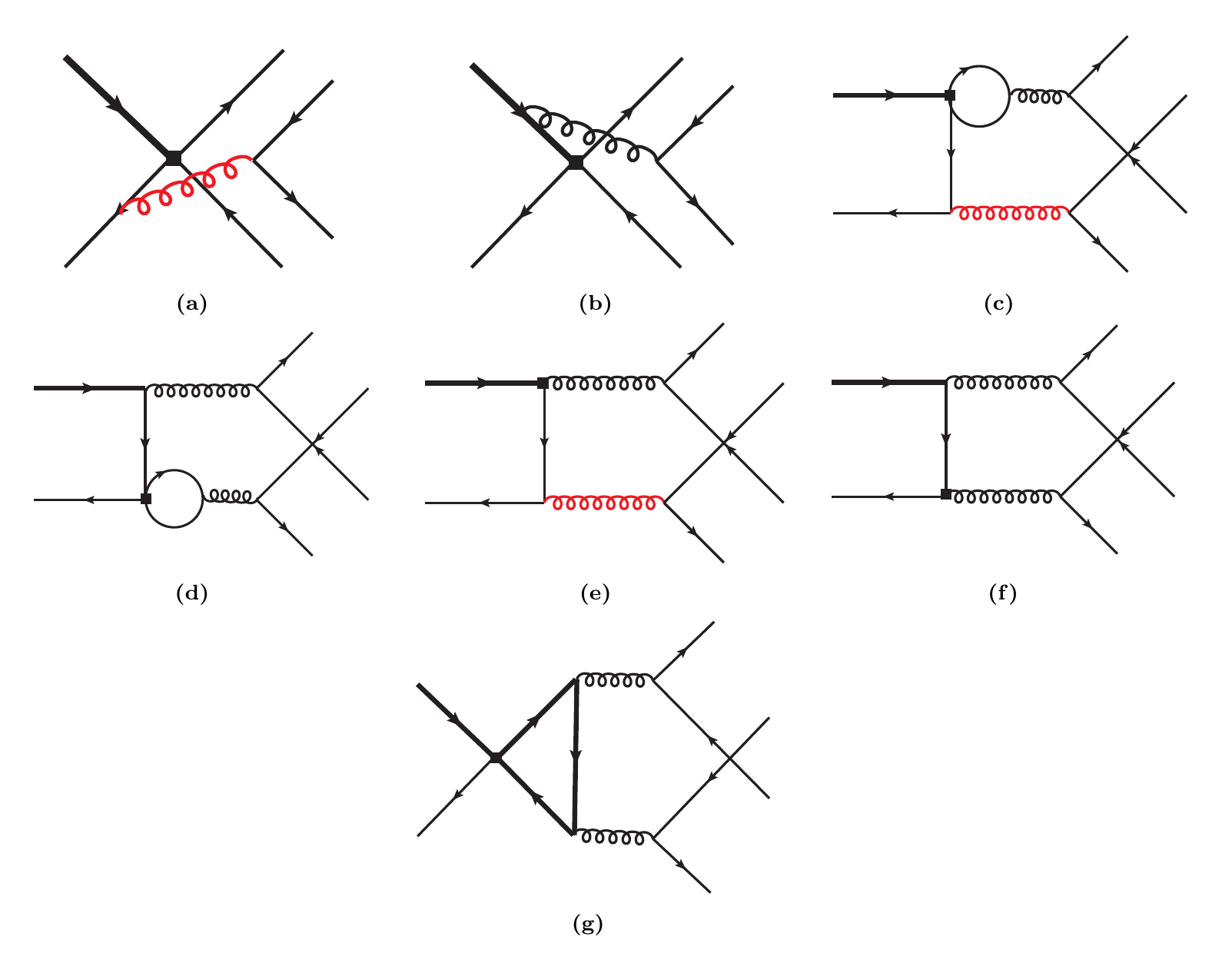}
\caption{Sample Feynman diagrams for the weak annihilation $\bar B_q \to M_1 M_2$ decays
at LO and NLO. The gluons marked with red colour can
carry either the hard or hard-collinear momentum, while the remaining gluons can only possess the hard momentum.
Symmetric diagrams that follow from (c)-(g) by exchanging the two off-shell gluons
are not displayed.}
\label{fig: annihilation diagrams at LO and at NLO}
\end{center}
\end{figure}

It proves convenient to cast the exclusive transition amplitude governing
the  weak-annihilation  bottom-meson decay in the form
\begin{eqnarray}
&& \bar {\cal A} (\bar B_q \to M_1 M_2)
= - \langle M_1 (p_1)  \, M_2 (p_2)  | {\cal H}_{\rm eff}  | \bar  B_q (p_B) \rangle
\nonumber \\
&& =  \mp \, i \, {4 \, G_F \over \sqrt{2}} \,  f_B \, f_{M_1} \, f_{M_2}
\, \sum_{p=u, c} V_{p b} \, V_{p q}^{\ast} \,
\left ( \pi \, \alpha_s \right )   \, {C_F \over N_c^2} \,
\nonumber \\
&&  \hspace{0.5 cm}
\times
\left [  {\cal T}^{p, \, (0)}
+ \left ( {\alpha_s \over 4 \, \pi} \right )  \, {\cal T}^{p, \, (1)}
+  {\cal O}(\alpha_s^2) \right ] \,,
\hspace{0.5 cm}
\label{Amplitude: master formula}
\end{eqnarray}
where  the upper sign in (\ref{Amplitude: master formula}) applies for
$M_1 \, M_2 = PP, \, V_{L} V_{L}$, while the lower sign when $M_1 \, M_2 = PV, \, VP$.
We leave out the discussion on final states consisting of two transversely
polarized vector mesons, bearing in mind that the resulting helicity amplitudes
are power-suppressed 
when compared with the counterpart longitudinal polarization amplitude \cite{Beneke:2006hg}.
The appearing LO quantities ${\cal T}^{p, \, (0)}$ can be  determined
by evaluating the first two Feynman diagrams in
Figure \ref{fig: annihilation diagrams at LO and at NLO},
where we also display the relevant diagrams generating
the dynamically enhanced contributions to the NLO pieces ${\cal T}^{p, \, (1)}$.
It remains important to remark that the additional LO annihilation diagrams
with the gluon radiation  off the final-state partons
(not included in Figure \ref{fig: annihilation diagrams at LO and at NLO})
lead to the vanishing effects for the pure annihilation channels
in the limit of symmetric distribution amplitudes and  under the assumption of
the ${\rm SU(3)}$ flavour symmetry as noted  in \cite{Beneke:2000ry}.

Taking advantage of  the flavour-decomposition strategy \cite{Beneke:2003zv}
allows us to write down
\begin{eqnarray}
{\cal T}^{p, \, (0)} &=& \delta_{p u} \,
{\cal C}_{M_1 M_2}^{(1)} \,  {\cal B}_{1}(M_1 M_2)
+  \, {\cal C}_{M_1 M_2}^{p, \,  (4)} \,  {\cal B}_{4}(M_1 M_2)
\nonumber \\
&& + \, {\cal C}_{M_1 M_2}^{p, \, (4, {\rm EW})} \,  {\cal B}_{4, \rm {EW}}(M_1 M_2) \,.
\end{eqnarray}
The prefactors ${\cal C}_{M_1 M_2}^{(p), \, (i, \, (\rm EW))}$ collect the Clebsh-Gordan coefficients
from the flavour structures of the $\bar B_q$-meson as well as
the final-state mesons and they further absorb the electric-charge coefficients from $P_{i Q}$.
The phenomenologically dominating effects of  ${\cal T}^{p, \, (1)}$
can be categorized in terms of their topological structures
\begin{eqnarray}
{\cal T}^{p, \, (1)} & \supset & \sum_{i=1}^{6} C_i \, {\cal P}_i^{p, (1)}(M_1 M_2)
+ C_{8}^{\rm eff} \, {\cal P}_{8 g}^{(1)}(M_1 M_2)
\nonumber \\
&& + \sum_{i=1}^{6} C_i \, {\cal I}_i^{p, (1)}(M_1 M_2) \,,
\label{master formula of the NLO amplitude}
\end{eqnarray}
with $C_{8}^{\rm eff} =  C_{8 g} + C_3 - C_4 /6 + 20 \, C_5 - 10 C_6 /3$ \cite{Beneke:2001at}.
We will dedicate the next section to the analytical computation  of
the flavour amplitudes ${\cal B}_{i, ({\rm EW})}$
and the NLO building blocks ${\cal P}_{1...6, 8g}^{p, (1)}$
and ${\cal I}_i^{p, (1)}$ within QCD factorization.

\section{QCD factorization for decay amplitudes}
%

Establishing the factorization formulae for the hadronic quantities
${\cal B}_{i, ({\rm EW})}$ can be customarily achieved by investigating
the appropriate partonic amplitudes of the tree diagrams (a) and (b)
in Figure \ref{fig: annihilation diagrams at LO and at NLO} in the leading-twist approximation.
The yielding contribution due to the gluon emission from the bottom quark (i.e., the LO diagram (b))
has been demonstrated to be calculable self-consistently at twist-two order \cite{Chernyak:1981zz,Beneke:2000ry}.
We are then led to anatomize the infrared behaviour of the effective matrix element
represented by the first diagram  in Figure \ref{fig: annihilation diagrams at LO and at NLO}
(with an insertion of $P_2^{u}$ for the illustration purpose)
\begin{eqnarray}
\langle P_2^{u} \rangle^{(0)}_{(a)}
&=& \left ( \pi \, \alpha_s \right ) \,  \frac{C_F}{N_c^2} \,
\int_0^{\infty} d \omega \, \int_0^1 d x \, \int_0^1 d y \,
\nonumber \\
&& {\rm Tr} \bigg \{ {\cal M}^{B}(v, \omega) \,
\left [ \gamma_{\perp \nu} \, \left ( \slashed{p}_1 + \slashed{q}_2 - \slashed{k} \right )
\gamma_{\mu} \left ( 1 - \gamma_5 \right ) \right ] \,
\nonumber \\
&& \hspace{0.5 cm} {\cal M}^{M_1}(p, y) \, \left [ \gamma_{\perp}^{\nu} \right ] \,
{\cal M}^{M_2}(q, x) \, \left [ \gamma^{\mu}  \left ( 1 - \gamma_5 \right ) \right ]  \bigg \}
\nonumber \\
&& \times \,  \frac{1}{ [(p_1 + q_2 - k)^2 + i \epsilon] [(p_1 + q_2)^2 + i \epsilon] } \,,
\label{result of the diagram a}
\end{eqnarray}
where the explicit expressions of the momentum-space projection operators
${\cal M}^{B}, \, {\cal M}^{M_1}$ and ${\cal M}^{M_2}$   have been derived
in \cite{Beneke:2000wa,Beneke:2001ev,Beneke:2002bs,Beneke:2006hg}
(see \cite{Grozin:1996pq} for the original discussion on $B$-meson distribution amplitudes
in heavy quark effective theory (HQET) and  \cite{Bell:2013tfa} for an alternative construction
of the on-shell light-cone projectors).
Here we have assigned the four-momenta $p_{1, 2}$  ($q_{1, 2}$)
to the quark and antiquark in the energetic meson $M_1$ ($M_2$).
The dimensionless variable $y$ ($x$) refers to
the momentum fraction carried by the quark field
in the composite  $M_1$ ($M_2$) system in the (anti)-collinear limit.
In addition,  the longitudinal momentum component $\omega$ for
the light partonic constituent of the $B$-meson state can be defined by
the kinematic relation  $p_1 \cdot k = \bar n \cdot p_1 \, n \cdot k / 2
\equiv \bar n \cdot p_1 \, \omega  / 2$ with the collinear-$p_1$ approximation,
where the two light-cone vectors $n_{\mu}$ and $\bar n_{\mu}$
fulfill the constraints $n^2=\bar n^2=0$ and $n \cdot \bar n=2$.
It is then convenient to simplify the quark/gluon propagators
in the last line of (\ref{result of the diagram a}) to
\begin{eqnarray}
\frac{1}{y \, m_B^2 \, \left [ \bar x - {\omega / m_B} + i \epsilon \right ]} \,
\frac{1}{y \, \bar x \,  m_B^2 + i \epsilon } \,,
\label{quark-gluon propagator}
\end{eqnarray}
by applying the kinematic properties of $\bar B_q \to M_1 M_2$,
the end-point behaviours of the twist-two distribution amplitudes
for the final-state mesons \cite{Braun:1989iv,Ball:1998sk},
and the momentum-fraction scaling rules of the Dirac trace
in the second and third lines of (\ref{result of the diagram a}).

We can readily verify that the obtained result (\ref{quark-gluon propagator})
applies for both the generic $\bar x \sim {\cal O} (1)$  and
the end-point region $\bar x \sim {\cal O} (\Lambda_{\rm QCD} / m_B)$,
which  correspond to the anti-collinear and
anti-soft-collinear \cite{Becher:2003qh,Becher:2015hka} momentum $q_2$.
This  observation is  in analogy to the smooth interpolation of the $A$-type contribution
to $\bar B_q \to \gamma \ell \bar \ell$ described by
the ${\rm SCET_I}$ factorization formula (3.14) in \cite{Beneke:2020fot}.
In the anti-collinear $q_2$ case, the resulting convolution integrals
in (\ref{result of the diagram a})  become
$\int_0^{\infty} d \omega \, \phi_B^{+} (\omega)  \,
\int_0^1 d x \, \int_0^1 d y \, {\phi_{M_2}(x) \, \phi_{M_1}(y) / (\bar x^2 \,  y)}$
by dropping out the subleading  term $\omega / m_B$  in (\ref{quark-gluon propagator}),
which reproduce exactly the obtained expression in \cite{Beneke:2001ev}
and develop the rapidity divergence thus spoiling the soft-collinear factorization
at leading twist.
The appearance of such singularity as $\bar x \to 0$ implies that
both the hard and  hard-collinear gluon exchanges
(labelled with the red colour in Figure \ref{fig: annihilation diagrams at LO and at NLO})
will bring about the leading-power contribution to the flavour amplitude  ${\cal B}_{1}$,
in spite of the phase-space suppression for the size of the hard-collinear region,
which is compensated by the QCD enhancement from the hard-collinear quark and gluon propagators.

Having in mind the essential importance of including the hard and hard-collinear gluon
exchange effects simultaneously, we proceed to write down the  convolution integrals
entering the factorized matrix  element $\langle P_2^{u} \rangle^{(0)}_{(a)}$
\begin{eqnarray}
\hspace{0.6 cm}
{\cal G}_{{\cal B}_{1}} &\equiv& \int_0^{\infty} d \omega  \,  \phi_B^{+} (\omega)
\int_0^1 d x   \, \phi_{M_2}(x)    \int_0^1 d y  \, \phi_{M_1}(y)
\nonumber \\
&&  \frac{1}
{\bar x  \,  y \,\left ( \bar x - {\omega / m_B} + i \epsilon \right )} \,,
\label{improved factorization formula}
\end{eqnarray}
which can be  reduced to a rather compact form
${\cal G}_{{\cal B}_{1}}  \approx
18 \,\left [ \left (\ln {m_B \over \lambda_B}  + \gamma_{\rm E} - 2 \right ) - i \, \pi \right ]$
in the leading-power approximation by employing the Grozin-Neubert model \cite{Grozin:1996pq} for $\phi_B^{+} (\omega)$
and the asymptotic twist-two distribution amplitudes $\phi_{M_2}(x)$ and $ \phi_{M_1}(y)$.
The dimensionful quantity $\lambda_B^{-1}$ represents the inverse moment of
the above HQET distribution amplitude \cite{Grozin:1996pq,Korchemsky:1999qb,Beneke:2000wa}
and serves as an indispensable ingredient for the theory description of
a wide variety of exclusive bottom-meson decays
\cite{Korchemsky:1999qb,Descotes-Genon:2002crx,Bosch:2003fc,Beneke:2011nf,Wang:2016qii,Wang:2018wfj,
Beneke:2018wjp,Shen:2020hfq,Beneke:2020fot,Beneke:2021rjf,Wang:2021yrr}.
It is worthwhile to stress that we do not aim at implementing QCD resummation
of {\it the  perturbatively  generated logarithms} of ${m_B / \lambda_B}$,
which are nevertheless not numerically significant for the realistic $\bar B_q$-meson mass.
The resulting  factorization formula (\ref{improved factorization formula})
provides a dynamical interpretation of the non-perturbative object $X_A^{i}$
introduced in \cite{Beneke:2001ev,Beneke:2003zv}
(with the superscript ``{\it i}" characterizing the
 gluon emission from the initial-state quarks) with
$X_A^{i} \approx \left [ 1  + \frac{\left ( \gamma_{\rm E} - 1 \right )  - i \, \pi }
{\ln {m_B \over \lambda_B}} \right ]  \, \ln {m_B \over \lambda_B}$
explicitly,
which can be matched onto the very parametrization
$X_{A, \, \rm {BBNS}}^{i}   = (1 + \varrho_A \, e^{i \, \varphi_A})  \,
\ln {m_B \over \Lambda_h}$
suggested in \cite{Beneke:2001ev,Beneke:2003zv}
by setting $\Lambda_h = \lambda_B$ and $(\varrho_A, \varphi_A)= \left \{ (0.97, \, -97 \degree) \right. $,
$(1.17, \, -97 \degree)$, $\left. (1.34, \, -97 \degree) \right \}$
for $\lambda_B =  \{ 200, \, 350, \, 500 \} \,  {\rm MeV}$, respectively.

\begin{widetext}
We are now ready to present the factorized expressions for the tree-level flavour amplitudes
\begin{eqnarray}
{\cal B}_1 &=& {1 \over 4} \, \left ( C_2 - { C_1  \over 2 \, N_c} \right ) \, \widehat {\cal G}_{{\cal B}_{1}},
\,\,
{\cal B}_{4} = {1 \over 8} \, \left [ \left ( C_4 + 16 \, C_6 \right ) \, \widehat {\cal G}_{{\cal B}_{1}}
+   \left ( C_4 + 4 \, C_6 \right ) \, \, \widehat {\cal G}_{{\cal B}_{4}} \right ],
\,\,
{\cal B}_{4, {\rm EW}} =  {\cal B}_{4} \left (  C_4 \to C_{4 Q}, \, C_6 \to C_{6 Q} \right ),
\hspace{0.6 cm}
\end{eqnarray}
where
$\widehat {\cal G}_{{\cal B}_{1}} = {\cal G}_{{\cal B}_{1}}
+  \int_0^1 d x  \, \int_0^1 d y \,
{ \phi_{M_2}(x) \, \phi_{M_1}(y) / (y \, (1- x \, \bar y))}$
and $\widehat {\cal G}_{{\cal B}_{4}}$ can be obtained from $\widehat {\cal G}_{{\cal B}_{1}}$
by performing the replacement $x \leftrightarrow \bar y$ for the counterpart short-distance matching coefficient.
Along the same vein we can derive the relevant results for the NLO topological amplitudes
\begin{eqnarray}
\sum_{i=1}^{6} C_i \, {\cal P}_i^{p, (1)} &=&
 \left ( C_2 - { C_1  \over 2 \, N_c} \right ) \, H_{\rm 1}(m_p)
 +  \left [ \left ( C_3 + 16 \, C_5 \right ) - {1 \over 2 \, N_c} \, (C_4 + 16 \, C_6 ) \right ] \,
\left [  H_{1}(m_b) + H_{1}(0) \right ]
\nonumber \\
&& \hspace{0.2 cm} + \left (C_4 + 10 \, C_6 \right ) \, \left [  H_{1}(m_b) + H_{1}(m_c) + 3 \, H_{1}(0) \right ]
- \left [ 5 \, C_4 - 8 \, C_5
+ 4 \, \left ( {1 \over N_c} + 5 \right ) C_6 \right ]  H_{2} \,,
\nonumber \\
C_{8}^{\rm eff}  \, {\cal P}_{8 g}^{(1)} &=&  C_{8}^{\rm eff}  \, H_{3} \,,
\qquad
\sum_{i=1}^{6} C_i \, {\cal I}_i^{p, (1)}  =
\left [ \left ( C_3 + 4 \, C_5 \right ) + C_F \, (C_4 + 4 \, C_6 ) \right ] \, H_{4} \,,
\label{NLO amplitudes}
\end{eqnarray}
\end{widetext}
where 
the explicit expressions of the  $H_i$ functions  are
presented in the Supplemental Material.
In contrast to QCD factorization for the penguin contributions $H_{1,2,3}$,
the one-loop building blocks ${\cal I}_i^{p, (1)}$ from the triangle diagram
in Figure  \ref{fig: annihilation diagrams at LO and at NLO} (g)
(plus the one with the two virtual gluons exchanged) are insensitive to
the (anti)-hard-collinear dynamics at leading power,
thus ensuring the disappearance of  the HQET $B$-meson  distribution amplitudes
in the  factorization formula for $H_4$.
Adopting the ans\"{a}tz for the twist-two light-meson distribution amplitudes $\phi_{M_1}(x) = \phi_{M_2}(x)$,
the peculiar linear combinations of Wilson coefficients $C_i$
in the yielding result (\ref{NLO amplitudes}) for 
$\sum \limits_{i=1}^{6} C_i \, {\cal I}_i^{p, (1)}$ coincide with the emergent patterns
for the annihilation-type contributions to the axial-vector form factors of
$\bar B_{q} \to \gamma \ell \bar \ell, \, \gamma \gamma$ \cite{Beneke:2020fot,Shen:2020hfq}.
This interesting observation, on the one hand,  stems from
the Bose-Einstein statistics for the transition matrix element
\begin{eqnarray}
&& \langle g^{\ast}(p_g, \alpha) \,  g^{\ast}(\tilde{p}_g, \beta) |{\cal H}_{\rm eff}  | \bar B_{q} \rangle
\nonumber \\
&& = i \, \epsilon_{\alpha \beta p_g \tilde{p}_g} \, F_V(p_g^2, \, \tilde{p}_g^2)
+ g_{\alpha \beta}^{\perp} \, \, F_A(p_g^2, \, \tilde{p}_g^2)
\label{B to gg matrix element}
\end{eqnarray}
with the two transversely polarized gluons,
which leads  to the form-factor relations $F_{V}(p_g^2, \, \tilde{p}_g^2)= - F_{V}(\tilde{p}_g^2, \, p_g^2)$
and $F_A(p_g^2, \, \tilde{p}_g^2)= F_A(\tilde{p}_g^2, \, p_g^2)$.
On the other hand, this can be attributed to the symmetry constraints of the QCD matrix element
\begin{eqnarray}
&& \langle M_1 (p) \, M_2 (q) | g^{\ast}(p_g, \alpha) \,  g^{\ast}(\tilde{p}_g, \beta) \rangle
\nonumber \\
&& =  \left\{
\begin{array}{l}
g_{\alpha \beta}^{\perp} \, \, {\cal S}_{\|}(M_1 M_2)
\hspace{1.2 cm}  {\rm for} \,\, M_1 M_2 = PP, \, V_L V_L  \,,
\\
i \, \epsilon_{\alpha \beta p q} \,\, {\cal S}_{\perp} (M_1 M_2)
\qquad {\rm for} \,\, M_1 M_2 = PV, \, VP  \,,
\end{array}
\right. \hspace{0.6 cm}
\label{gg to M1 M2 matrix element}
\end{eqnarray}
which guarantee the transversity amplitudes ${\cal S}_{\|, \perp}$ proportional to
the products of the twist-two distribution amplitudes $\phi_{M_2}(x) \, \phi_{M_1}(y)$.
Combining together the  requirements of the hadronic matrix elements
(\ref{B to gg matrix element}) and (\ref{gg to M1 M2 matrix element})  as discussed above,
we are then led to conclude that only the axial-vector form factor $F_A$ will be  in demand
when evaluating the NLO  quantities displayed in (\ref{master formula of the NLO amplitude}).
It is also apparent that the observed perturbative enhancement 
due to the penguin contractions of $P_{1, 2}^{c}$
does not apply to the charmless  $\bar B_q \to P V$ decays,
which will therefore not be discussed in the subsequent numerical analysis.

\section{Phenomenological implications }
%

We are now equipped to explore the phenomenological significance of the factorized expressions
for the hadronic quantities ${\cal T}^{p, \, (0)}$ and ${\cal T}^{p, \, (1)}$
dictating  the exclusive $\bar B_q \to M_1 M_2$ decay amplitude (\ref{Amplitude: master formula}).
To this end, we will employ the three-parameter ans\"{a}tz of the leading-twist
bottom-meson distribution amplitude as proposed in \cite{Beneke:2018wjp}
(see \cite{Kawamura:2001jm,Kawamura:2001bp,Huang:2005kk,Wu:2013lga,
Khodjamirian:2010vf,Wang:2017jow,Braun:2017liq,Lu:2018cfc,Gao:2019lta,Gao:2021sav}
for additional discussions on the higher-twist distribution amplitudes)
$\phi_B^{+}(\omega, \, \mu_0)
=  {\Gamma(\beta) \over \Gamma(\alpha)} \,  U \left (\beta-\alpha, 3 - \alpha, {\omega \over \omega_0} \right )
{\omega \over \omega_0^2}  {\rm exp} \left ( - {\omega \over \omega_0} \right )$
at the reference scale $\mu_0= 1 \, {\rm GeV}$,
where $U(a, b, z)$ represents the confluent hypergeometric function of the second kind
and the allowed intervals of the shape parameters $\omega_0$, $\alpha$ and $\beta$
are determined by reproducing the numerical values of the inverse-logarithmic moments
\cite{Beneke:2020fot,Shen:2020hfq} $\lambda_{B_d} = 350 \pm 150 \, {\rm MeV}$,
$\lambda_{B_s} = 400 \pm 150 \, {\rm MeV}$, $\widehat{\sigma}_{B_{d, s}}^{(1)} = 0.0 \pm 0.7$,
$\widehat{\sigma}_{B_{d, s}}^{(2)} = 0.0 \pm 6.0$.
We proceed to  adopt the lattice predictions for the pseudoscalar-meson decay constants
($f_{B_d}$, $f_{B_s}$, $f_{\pi}$, $f_{K}$) as summarized in \cite{Aoki:2021kgd}
and take the improved extractions of the  decay constants of the longitudinally
polarized vector mesons ($f_{\rho}$, $f_{\omega}$, $f_{\phi}$) from \cite{Bharucha:2015bzk}.
Furthermore,  we will truncate the Gegenbauer expansions of the twist-two light-meson distribution amplitudes
at the next-to-next-to-leading conformal spin accuracy and take advantage of the non-perturbative determinations
of the two lowest moments $a_{1, 2}(\mu_0)$ from \cite{RQCD:2019osh} for the energetic pseudoscalar mesons
and from \cite{Bharucha:2015bzk} for the light vector mesons.
The default values and uncertainties for the remaining SM parameters follow  Table 1 of \cite{Shen:2020hfq}.

\begin{table}
\centering
\renewcommand{\arraystretch}{2.0}
\resizebox{\columnwidth}{!}{
\begin{tabular}{c|c|c|c}
  \hline
  \hline
   & \,\, $\nu=m_b/2$  \,\, &  \,\, $\nu=m_b$   \,\, &  \,\, $\nu=2 \, m_b$   \,\, \\
  \hline
 ${\cal T}^{u, \, (0)}$ & $2.03 - 17.7 \, i$ & $2.00 - 17.4 \, i$  & $1.97 - 17.2 \, i$ \\
 ${\cal T}^{c, \, (0)}$ & $-0.38 + 3.35  \, i$ & $-0.25 + 2.16 \, i$ & $-0.16 + 1.42 \, i$  \\
 \,\, ${\alpha_s / (4 \, \pi)}  \,\, {\cal T}^{u, \, (1)}$  \,\, & $-2.29 - 3.63 \, i$ &  $-2.41 - 3.61 \, i$ & $-2.47 - 3.57 \, i$ \\
 $ \,\,  {\alpha_s / (4 \, \pi)}  \,\, {\cal T}^{c, \, (1)}$  \,\, & $-1.65 - 2.34 \, i$ & $-1.82 - 2.40 \, i$ & $-1.91 - 2.43 \, i$ \\
  \hline
  \hline
\end{tabular}
}
\renewcommand{\arraystretch}{1.0}
\caption{Numerical predictions of the dynamical quantities ${\cal T}^{p, \, (0)}$
and ${\cal T}^{p, \, (1)}$  ($p=u, \, c$) for $\bar B_s \to \pi^{+}  \pi^{-}$
with three distinct renormalization scales $\nu$.}
\label{table for the predicted amplitudes}
\end{table}

In order to develop a transparent understanding of the numerical feature
for the  identified enhancement mechanism,
we present in Table \ref{table for the predicted amplitudes} the obtained results of 
${\cal T}^{p, \, (0)}$ and  ${\cal T}^{p, \, (1)}$ for the weak annihilation decay $\bar B_s \to \pi^{+}  \pi^{-}$
with three distinct values of the renormalization  scale
$\nu=m_b/2, \, m_b, \, 2 \, m_b$,
while employing the central values of the remaining theory inputs.
Interestingly, the dominant NLO contribution from the charm-loop diagrams
displayed in Figure  \ref{fig: annihilation diagrams at LO and at NLO} (c) and (d)
(plus the two diagrams due to the crossing symmetry) will result in the significant increase of
the real part of ${\cal T}^{c, \, (0)}$ in magnitude
and simultaneously generates the considerable cancellation of its imaginary part,
keeping in mind the  CKM hierarchy $V_{ub} V_{us}^{\ast} / V_{cb} V_{cs}^{\ast} = 0.021 \, e^{- 1.20 \, i}$.
In consequence,  the perturbative rescattering effect under discussion provides an important source of
the strong phases of the weak-annihilation non-leptonic $\bar B_q$-meson decay amplitudes,
which can then be probed feasibly  by investigating the CP violating observables.

\begin{table}
\centering
\renewcommand{\arraystretch}{2.0}
\resizebox{\columnwidth}{!}{
\begin{tabular}{c|c|c}
  \hline
  \hline
   & \,\, ${\cal A}_{\rm CP}^{\rm dir}$  \,\, &  \,\, ${\cal A}_{\rm CP}^{\rm mix}$   \,\,  \\
  \hline
 \,\, $\bar B_s \to \pi^{+} \, \pi^{-}, \, \pi^{0} \, \pi^{0}$  \,\, & $-36.3^{+8.2}_{-1.3}$ ($0.0 \pm 0.0$)
 & $-4.2^{+21.4}_{-9.0}$ ($35.9^{+15.6}_{-11.2}$)   \\
 $\bar B_s \to \rho^{+}_{L} \, \rho^{-}_{L} , \, \rho^{0}_{L}  \, \rho^{0}_{L} $ & $-36.3^{+8.3}_{-1.8}$ ($0.0 \pm 0.0$)
 &  $-4.3^{+21.5}_{-9.0}$ ($35.9^{+15.6}_{-11.2}$) \\
 $\bar B_s \to \omega_{L}  \, \omega_{L} $ &  $-36.3^{+8.3}_{-3.1}$ ($0.0 \pm 0.0$)
 &  $-3.8^{+21.8}_{-9.7}$ ($35.9^{+15.6}_{-11.2}$) \\
 $\bar B_s \to \rho_{L}  \, \omega_{L} $ & $0.0 \pm 0.0$ ($0.0 \pm 0.0$ )
 &  $-71.0^{+6.3}_{-5.4}$ ($-71.0^{+6.3}_{-5.4}$) \\
 \hline
 $\bar B_d \to K^{+} \, K^{-}$ &  $39.0^{+3.2}_{-5.6}$ ($0.0 \pm 0.0$)
 &  $-2.2^{+19.1}_{-26.4}$ ($-47.0^{+15.7}_{-18.8}$) \\
 $\bar B_d \to K^{\ast +}_{L}  \, K^{\ast -}_{L} $ & $39.6^{+4.9}_{-6.7}$ ($0.0 \pm 0.0$)
 &  $-1.4^{+19.7}_{-26.9}$ ($-47.0^{+15.7}_{-18.8}$) \\
 $\bar B_d \to \phi_{L}  \, \phi_{L} $ & $38.3^{+11.4}_{-15.8}$ ($0.0 \pm 0.0$)
 & $27.8^{+5.7}_{-25.9}$  ($0.0 \pm 0.0$) \\
  \hline
  \hline
\end{tabular}
}
\renewcommand{\arraystretch}{1.0}
\caption{Theory predictions of the CP asymmetries ${\cal A}_{\rm CP}^{\rm dir}$
and  ${\cal A}_{\rm CP}^{\rm mix}$ (in unites of $10^{-2}$)
for the weak annihilation $\bar B_q \to P P, \, V_L V_L$ decay processes,
where we have included in parentheses the corresponding LO QCD results for a comparison.}
\label{table for the predicted CP asymmetries}
\end{table}

It is straightforward to derive the time-dependent CP asymmetries for the neutral $\bar B_q$-meson
decaying into CP eigenstates
\begin{eqnarray}
{\cal A}_{\rm CP}(t) &=& \frac{\Gamma (\bar B_q \to M_1 M_2)-\Gamma (B_q \to M_1 M_2)}
{\Gamma (\bar B_q \to M_1 M_2)+\Gamma (B_q \to M_1 M_2)}
\nonumber \\
&=& - \frac{{\cal A}_{\rm CP}^{\rm dir} \, \cos (\Delta m_q \, t) + {\cal A}_{\rm CP}^{\rm mix} \, \sin (\Delta m_q \, t)}
{\cosh(\Delta \Gamma_q \, t /2) + {\cal A}_{\Delta \Gamma} \,  \sinh(\Delta \Gamma_q \, t /2)} \,,
\hspace{0.5 cm}
\end{eqnarray}
where the general definitions of ${\cal A}_{\rm CP}^{\rm dir}$, ${\cal A}_{\rm CP}^{\rm mix}$
and ${\cal A}_{\Delta \Gamma}$ can be found, for instance, in
\cite{ParticleDataGroup:2020ssz,Nierste:2009wg,Proceedings:2001rdi},
satisfying the exact algebra relation
$|{\cal A}_{\rm CP}^{\rm dir}|^2 + |{\cal A}_{\rm CP}^{\rm mix}|^2 + |{\cal A}_{\Delta \Gamma}|^2 = 1$.
The resulting predictions for the two independent CP asymmetries
displayed in Table \ref{table for the predicted CP asymmetries}
evidently reveal the striking numerical impacts of the enhanced NLO QCD correction to
the weak annihilation amplitudes described in this Letter,
without which the direct CP violation for  the entire pure annihilation channels
would vanish due to the absence of at least two separate terms with different
strong phases involved in $\bar {\cal A} (\bar B_q \to M_1 M_2)$.
By contrast, the time-dependent CP asymmetries of $\bar B_s \to \rho_{L}  \, \omega_{L} $
turn out to be insensitive to this particular perturbative contribution in question,
on account of the isopsin non-conservation of the QCD scattering amplitude
$ \langle \rho_{L}(p)  \, \omega_{L}(q) | g^{\ast}(p_g, \alpha) \,  g^{\ast}(\tilde{p}_g, \beta) \rangle$.
We further mention in passing that the newly computed NLO quantities ${\cal T}^{p, \, (1)}$
generate the subdominant corrections to the CP-averaged branching fractions
of the pure annihilation decay channels (numerically at the level of ${\cal O} (10 \, \%)$).

%
\section{Conclusions}
%

To summarize, we have endeavored to achieve the analytical regularization of
end-point divergences in the factorization formulae
of the peculiar weak annihilation non-leptonic bottom-meson decay amplitudes,
due to the  gluon radiation off the initial-state partons,
by adding the missing hard-collinear contribution on top of the hard gluon exchange effect,
both of which led to the leading power contributions in the heavy quark expansion.
Applying the improved factorization formalism, we then identified the novel perturbative mechanism
generated by the  penguin contractions of the current-current operators,
yielding the significant numerical impacts on the strong phases of the pure annihilation amplitudes.
Our results are of importance for enhancing the predictive power of
the perturbative factorization approach in addressing the entire spectrum of
CP violating observables from   exclusive heavy hadron decays,
accessible at the LHCb and Belle II experiments.

%
\begin{acknowledgments}
\section*{Acknowledgements}

We would like to thank Martin Beneke for interesting discussions and comments on the manuscript.
The research of C.D.L. is supported by  the  National Natural Science Foundation of China  with
Grant No. 11521505 and 12070131001 as well as the National Key Research and Development Program of China
under Contract No. 2020YFA0406400.
C.W. is supported in part by the National Natural Science Foundation of China
with Grant No. 12105112 and  the Natural Science Foundation of
Jiangsu Education Committee with Grant No. 21KJB140027.
The research of Y.L.S. is supported by the  National Natural Science Foundation of China  with
Grant No. 12175218 and the Natural Science Foundation of Shandong with Grant No.  ZR2020MA093.
Y.M.W. acknowledges support from the  National Natural Science Foundation of China  with
Grant No. 11735010 and 12075125, and the Natural Science Foundation of Tianjin
with Grant No. 19JCJQJC61100.

\end{acknowledgments}

%
\appendix
\begin{widetext}
\section{Supplemental Material: Analytic Expressions for the NLO Kernels }

We collect QCD factorization formulae for the primitive kernels
entering the NLO topological amplitudes (\ref{NLO amplitudes})
\begin{eqnarray}
&& H_{i \leq 3} =  {1 \over 12} \, \int_0^{\infty} d \omega  \,  \phi_B^{+} (\omega)  \int_0^1 d x   \, \phi_{M_2}(x)
\int_0^1 d y  \, \phi_{M_1}(y) \,  \left [ \left (  {h_i  \over \bar x  \,  y }  \,
\left ( \frac{1} {\bar x - {\omega / m_B} + i \epsilon } + {\eta_i \,  \bar x \over 1 - x \, \bar y} \right )
+  \{ x \leftrightarrow \bar y  \}  \right ) + \{ x \leftrightarrow y \}  \right ],
\nonumber \\
&& H_{4} =  2 \,  \int_0^1 d x   \, \phi_{M_2}(x)
\int_0^1 d y  \, \phi_{M_1}(y) \,  {1 \over \lambda} \,
\left [ h_4 \,  {\bar x + y \over \bar x \, y}
+ {1 \over x \, \bar x} \, \left ( {\pi \over \sqrt{3}}  - {\lambda \over 2 \, y \, \bar y}  \right )
+ \left (1 +  {\lambda \over 4 \, \bar x \, y}  \, \left (2 + {3 \over  x \, \bar y} \right ) \right ) \, \eta_4
+ \{ x \leftrightarrow y \}  \right ],
\nonumber \\
\end{eqnarray}
where for brevity  we have introduced the following perturbative functions
\begin{eqnarray}
h_{1} &=&  \ln {m_q^2 \over \mu^2}  + {1 \over 3} - {4 \,z_q  \over x \, \bar y}
+ 2 \, \left (1 +  {2 \,z_q \over x \bar y} \right ) \,
\sqrt{{4 \,z_q \over x \bar y} - 1} \, \arctan  {1 \over \sqrt{{4 \,z_q \over x \bar y} - 1}}  \,,
\qquad z_q = { m_q^2 \over m_B^2} \,,
\nonumber \\
h_{2} &=&  1, \qquad  h_{3}  = - {3 \over \bar y} \,, \qquad
h_{4} = - 2 \sqrt{{4  \over x \bar y} - 1} \,  \arctan  {1 \over \sqrt{{4  \over x \bar y} - 1}} \,,
\qquad  \eta_1 = \eta_2 =1,  \qquad \eta_3= - {y \over \bar x} \,,
\nonumber \\
\eta_4 &=& {1 \over \sqrt{\lambda}} \, \sum_{i=1}^{3} \left [ {\rm Li}_2 \left ( {\beta_i -1 \over  \alpha_i + \beta_i}  \right )
+ {\rm Li}_2 \left ( - {\beta_i - 1 \over  \alpha_i - \beta_i}  \right )
-  {\rm Li}_2 \left ( {\beta_i + 1 \over  \alpha_i + \beta_i}   \right )
-  {\rm Li}_2 \left (-  {\beta_i + 1 \over  \alpha_i - \beta_i}  \right ) \right ] \,,
\end{eqnarray}
with the K\"{a}ll\'{e}n function $\lambda \equiv \lambda (1, \bar x \, y, x \, \bar  y) =  (1-x-y)^2$
and
\begin{eqnarray}
\alpha_1 &=& \sqrt{1 +  {4 \over  - \bar x \, y - i \epsilon}}  \,,
\qquad
\alpha_2 = \sqrt{1 +  {4 \over  - x \, \bar y - i \epsilon}}  \,,
\qquad  \alpha_3 = \sqrt{5} \,,
\nonumber \\
\nonumber \\
\beta_1 &=&  { - \bar x \, y +  x \, \bar y  + 1 \over \sqrt{\lambda}}  \,,
\qquad
\beta_2 =  { - x \, \bar y +  1 + \bar x \, y \over \sqrt{\lambda}}  \,,
\qquad
\beta_3 =  { - 1 +  x \, \bar y + \bar x \, y \over \sqrt{\lambda}}  \,.
\end{eqnarray}

\end{widetext}

\bibliographystyle{apsrev4-1}

\bibliography{References}

\end{document}